# THE ADOPTION AND CHALLENGES OF ELECTRONIC VOTING TECHNOLOGIES WITHIN THE SOUTH AFRICAN CONTEXT

Mourine Achieng and Ephias Ruhode

Cape Peninsula University of Technology, Cape Town, South Africa

## Abstract

*Literature has shown that countries such as Brazil and India have successfully implemented electronic voting systems and other countries are at various piloting stages to address many challenges associated with manual paper based system such ascosts of physical ballot paper and other overheads, electoral delays, distribution of electoral materials, and general lack of confidence in the electoral process. It is in this context that this study explores how South African can leverage the opportunities that e-voting presents. Manual voting is often tedious, non-secure, and time-consuming, which leads us to think about using electronic facilities to make the process more efficient. This study proposes that the adoption of electronic voting technologies could perhaps mitigate some of these issues and challengesin the process improving the electoral process. The study used an on-line questionnaire which was administered to a broader group of voters and an in-depth semi-structured interview with the Independent Electoral Commission officials. The analysis is based on thematic analysis and diffusion of innovations theory is adopted as a theoretical lens of analysis. The findings reveal that relative advantage, compatibility and complexity would determine the intentions of South African voters and the Electoral Management Bodies (IEC) to adopt e-voting technologies. Moreover, the findings also reveal several other factorsthat could influence the adoption process. The study is limited to only voters in Cape Town and these voters were expected to have some access to the internet. The sample size limits the generalizability of the findings of this study.*

## Keywords

*Electronic voting technology, Diffusion of Innovation (DoI), E-democracy, &E-governance*

## Introduction

The technological development in South Africa has opened the possibilities of the use of ICTs in the democratic and governance processes. Garson views e-democracy as an umbrella term that covers many democratic activities carried out through electronic means and broadly defines e-democracy as "the use of ICT by governments to improve the efficiency, equity, and quality of democratic participation [1]."The applications of e-democracy include mechanisms to inform, consult, and broadly engage citizens through ICT use in the political process. These mechanisms are usually called "e-participation" or "e-engagement". Clift states that "e-democracy builds on e-governance and focuses on the actions and innovations enabled by ICTs combined with higher level of democratic motivation and intent [2]."Governance according toLai and Haleem is the system of leading and controlling the actions, affairs, policies and functions of a political unit, organization or nation[3].E-governance or electronic governance is using ICTs at various levels of the government and the public sector and beyond for the purpose of enhancing governance [4] - [6].

E-participation has been defined as the use of ICT supported platforms to facilitate the participation in the democracy and governance[7]. Voting forms an important part of democracy and for any country to be able to sustain democracy, voter participation is a key consideration. Rosenstone and Hansenemphasize that elections represent an important field to measure the ICT political use because of voting as the main participation channel[8]. Electronic voting (also known as e-voting) is a term encompassing several different types of voting technologies, including both electronic means of casting a vote and electronic means of counting votes. According to the ACE Electoral knowledge Network, countries like the USA, Brazil, and India have successfully implemented e-voting to address various challenges associated with the manual paper based electoral process [9]. This study analyses the challenges and prospects of adopting an e-voting system and how South Africa can leverage the opportunities it presents. The study also looks at the factors that could influence the adoption of electronic voting.The study looks at e-voting technologies against the manual paper based voting systems and does not look at a specific e-voting technology/system.

## 1. Background information

The South African constitution guarantees democracy in that; every citizen over the age 18 has the right to vote [10].South Africa held its first democratic election in 1994 and was run by the Independent Electoral Commission (IEC) and also included international observers who declared the electoral process free and fair.The IEC is a permanent body that was established in-terms of the Electoral Commission Act of 1996. It is independent from government, but reports to parliament.The IEC has been responsible for the implementation of the electoral system for elections in South Africa since the first democratic elections. A possible drawback to the current electoral system according to the SouthAfrica.info is the number of illiterate adults in South Africa[11].Another contributing drawback to the current electoral system is the level of poverty in South Africa. According to World Bank report South Africa 2009, 50% of South Africa's population still lived in under privileged conditions [12].Voters often complain of the distance they have to travel in order for them to reach polling stations. The logistical problems have often resulted into a decrease in voter registration and participation. People may not have the necessary funds to travel to an election polling station, which can be the reason for voter turnout decreasing in the South African elections. According toKersting in 1994, 84% of eligible voters cast their vote. In the 1999 election that figure declined to 63% and in 2004, the election had a 61% turnout [13].Another significant drawback is the difficulty in accessing some parts of the country, which makes distribution of ballot papers on Election Day very challenging.

According to theElectoral Institute of South Africa (EISA) report No. 12 2009, the cost of the physical ballot paper is also identified as drawback of the current paper based electoral system[14]. Another concern is confidence in the ballot forms. According to a survey done by Citizen Surveys in October and November 2008, South African citizens were concerned that the secrecy of their ballot forms could be compromised. A study was conducted on 2400South Africans revealing that 58% had confidence in the secrecy of their ballot forms. In a report delivered by the Electoral Institute of South Africa (EISA), the 2009 elections showed that large numbers of election officials did not have a clear understanding of the counting process which led to delay. They observed that some of the polling stations used one ballot box for both the national assembly and provincial legislatures. Also, the seals on some ballot boxes were not applied using the correct procedures [14].

Based on the background information it is clear that the current paper-based electoral process used in South Africa can be significantly improved to mitigate some of the challenges it is faced with. This study proposes that the adoption of electronic voting could perhaps drastically reduce some of these problems and in the process improving the electoral process. Some countries around the world are successfully implementing electronic voting systems to address many challenges associated with costs of physical ballot paper and other overheads, electoral delays, distribution of electoral materials, and general lack of confidence in the electoral process. South Africa however, has not leveraged the opportunities that e-voting presents. Manual voting is often tedious, non-secure, and time-consuming, which leads us to think about using electronic facilities to make the process more efficient.

## 2. E-governance in South Africa

The South African government has a number of e-governance initiatives that have been put into place since its inception in 1994. These initiatives are outlined in a project dubbed information communication year 2005. Kroukamp, lists some of the basic steps taken, including the installation of public information terminals for internet and e-mail access in certain rural centres as part of the joint public/private sector initiatives and the funding of computer centres in rural communities by companies such as Microsoft [15]. Some of these initiatives include South African Revenue Services (SARS) e-filing to facilitate the electronic submission of tax returns, the National Automated Archival Information Retrieval System (NAAIRS), provides extensive information and documentation about the national archive services to the public and to governments bodies.

The Department of Home Affair's National Identification System (HANIS) project, which has initiated an automated identification of database of fingerprints to combat crime and supply information for the purposes of policing[15]. According to Booz Allen Hamilton, when e-governance is used to its full potential, it can provide the convenience of electronic voting and encourages e-participation in public forums by all citizens[16].The State agency for Information Technology (SITA) is also an e-government initiative, SITA was established in 1999 to consolidate and coordinate the State's information technology resources in order to achieve cost savings through scale, increase delivery capabilities and enhance interoperability. SITA is committed to leveraging Information technology as a strategic resource for government, managing the IT procurement and delivery process to ensure that the government gets value for money, and using IT to support the delivery of e-government services to all citizens (www.sita.co.za).

## 3. Electronic voting technologies

Smith and Macintosh are of the view that a modern e-enabled system of democratic governance seems to require some sort of modernisation of the voting process, whether through e-counting methods or an e-voting system [17]. According to Macintoshthe most powerful symbol of a democracy is the citizen's involvement in the free and fair election of representatives to govern them [18]. Macintosh continues to state that voting is seen as the act that currently defines the relationships between citizens, governments and democracy. As such, e-voting takes on a powerful symbolic role in e-democracy. Qadah and Taha,define the term electronic voting as the "use of computers or computerized equipment to cast votes in an election[19]." They continue to emphasize that "e-voting aims at increasing participation, lowering the costs of running elections and improving the accuracy of the results [19]." According toSæbø, Rose and Skiftenes Flakelectronic voting is an activity within the e-participation field where e-participation actors (citizens, politicians, government institutions, voluntary organisations) conduct such an activity in the context of some factors (information availability, infrastructure, underlying technologies accessibility, etc.) which then result to certain effects like civic engagement, deliberative and democratic [20].

Electronic voting and counting technologies are increasingly being used around the world with India and Brazil taking the centre stage. Belgium and the Philippines also use electronic voting and counting technologies for their national elections. Countries such as Estonia, Norway, Pakistan, and United States are at various stages of piloting partially using electronic voting and counting technologies, including the use of internet voting. Brazil and India have successfully implemented e-voting to address various challenges associated with the manual paper based electoral process.

There are countries that are, however, moving away from e-voting technologies, for example the Netherlands in 2008 after several years of using e-voting machines, decertified all its machines and moved back to paper balloting. Germany likewise recently banned the use of e-voting machines it had been using. Ireland spent thousands of euros on e-voting machine that were only used in small piloting projects. In United States, the use of e-voting systems have always been controversial (the 2000 Florida elections) and has generated fierce debate between advocates and opponents of this technology. The worldwide experience of implementing e-voting is mixed with respect to adoption, non-adoption or adoption followed by abandonment.

There are various factors that could drive one nation towards an electronic voting or counting technologies which may not be present for another nation, or may indicate a different solution, for example, the challenges of moving paper ballots around large countries like Russia and Kazakhstan make the use of electronic voting technologies potentially beneficial on logistic grounds. The existence of a smart ID card with digital signature for the majority of the population in Estonia makes the use of Internet voting more feasible in Estonia.

The Philippines adopted an electronic counting solution to deal with issues related to fraud during the counting process. Factors that argue for or against the use of electronic voting or counting technologies in a particular country are specific to that country and will have many different sources – legal, cultural, political, logistical, environmental, etc.Technological developments in South Africa have opened the possibility of e-voting technologies and this clearly provides some opportunities and challenges. Svensson and Leenes argue that on the one hand, the electronic voting technology may help make voting more cost effective and more convenient for the voters and may even increase voter turnout. However, on the other hand, e-voting may introduce new risks and affect the electoral values such as secrecy of the vote and placing of voting as an observable institution in modern democracies [21].

Xenakis and Macintosh state that e-voting can play an important role in enhancing the voting process through increasing the low turnout among the youths who may not be satisfied with the traditional paper-based voting process [22]. According to Barker and Moon, e-voting can help in improving the participation of disabled people who are usually less likely to vote than individuals who have similar demographic characteristics [23]. This help comes through providing conveniences such as different text sizes, colour and audio voting; also e-voting can solve the mother tongue problem by allowing voting in different languages. All these benefits are in addition to the rapid and more accurate process of counting the votes that e-voting allows [24]. Although these benefits are attractive and can solve many problems of the paper-based voting system, the implementation of e-voting could present number of issues and challenges, including social, technical, political, legal and economic aspects. These aspects should be considered thoroughly before any decisions are made to unravel the hidden details that could affect the choice of suitable solutions and the overall success of the system.

## 4. Adoption Model

This study adopts Rogers's diffusion of innovation model as a theoretical framework [25].Rogers defines diffusion as "the process by which an innovation is communicated through certain channels over time among the members of a social society". Rogers further differentiates the adoption process from the diffusion process in that the diffusion process occurs within society, as a group process; whereas, the adoption process pertains to an individual. Rogers defines the adoption process as "the mental process through which an individual passes from first hearing an innovation to finally adopting the innovation [25].Diffusion theories are about the dynamic of change, in diffusion model the differences between or within a society are no more than a starting point. Diffusion models are not concerned with whether an innovation has been adopted, but when it is or will be adopted [26].Based on Lee et al. the diffusion of Innovation (DoI) model has emphasised that the attributes of new technology as key determinants of adoption[27]. This is one of the reasons the framework was chosen for this study.DoI theory according toRogers seeks to explain the process and factors that influence the adoption of new innovations; this is in line with the objective of this study, which is to explore the factors and challenges that could influence the adoption of electronic voting technology within the South African context.

The study uses three constructs from the DoI framework, relative Advantage, complexity and compatibility. These three innovation characteristics had the most consistent significant relationships to innovation adoption.According to Rogers individuals' perception of the attributes of an innovation and not the attributes as classified objectively by experts or change agents, affect the rate of adoption[25]. Rogers defines these attributes as follows: Relative advantage as "the degree to which an innovation is perceived as better than the idea it supersedes; Complexity, which is comparable to Technology Acceptance Model's perceived ease of use construct, as the degree to which an innovation is perceived by the potential adopter as being relatively difficult to use and understand and Compatibility as the degree to which an innovation is perceived to be compatible with existing values, beliefs, experiences and needs of potential adopters [25]."

Relative advantage was relevant for this research as the study wants to know how the voters in South Africa and the IEC officials compare the new technology (e-voting) to the current paper-based electoral process. For example the relative advantage will examine the research participants' point of view and the reasons they would be in favour or against electronic voting technologies.According to Taylor and Todd, compatibility is an important construct that can positively influence adoption[28].They give an example stating that "if the use of an innovation violates a cultural or social norm it is less likely to be adopted [28]." Complexity "is equivalent to ease of use (PEOU is the direct antonym of complexity) [29]-[31]." An example could be; finding out if voters perceive electronic voting to be easy to use or not, is similar to finding out if electronic voting could be complex to use or not; therefore the researcher has assumed it is the same.

## 5. Research Methodology

This research is an exploratory qualitative study with an interpretivist approach. Interpretivism is based on the observation that there are major differences between the natural world and the social world. Interpretivist researchers thus attempt to understand a phenomenon through accessing the meanings assigned to them by participants. The aim of interpretivism is to understand the individual experiences of those being studied, how they think and feel and how they act/re-act in their habitual contexts

The research's objectives was to explore how the adoption of e-voting would diffuse within South African context and also explore the factors that could influence the adoption of e-voting from the perspective of the citizens and the Independent Electoral Commission. To select the appropriate sample from the population was a challenging task for a number of reasons. Firstly, there is a high level of variation within the South African society, including a variation in age, culture, education, and relation to technology and secondly electronic voting technology is a new or nonexistence concept for South Africa citizens. The approach that this research used for selecting the sample was based on the notionthat when a new technology is introduced into a society, it is not expected to be adopted and used immediately by the entire population. On the contrary, the process of adoption passes through stages, starting with a small group of people who later after successfully adopting the technology encourages others to take up the new technology.

Based on this view, the researchselected two sampling unitswithin the Cape Town area. The first sample unitwas thevoters who were expected tobe more aware of technology and either eager to use technology or were currently using a technology similar to e-voting.The second sampling unit was the Independent Electoral Commission officials who are in charge of running the elections in South Africa. These two sampling units were assessed for their views on e-voting, their view on South Africa's readiness to adopt an e-voting system and how much they knew about e-voting technologies.

The study used an onlinesurvey questionnaire and in-depth semi-structured interview as the primary source of data. The questionnaire, which was designed in-line with guidelines for questionnaire design recommended by Babbie and Mouton, contained both open-ended and closed-ended questions and evolved around participants' view of e-voting technologies in general in comparison with the manual paper based voting systems currently used[32].

The questions asked were clear and simple to avoid double meanings. The questionnaire was then pre-tested with a small group of people before it was administered to the study population. This was done to see if the respondents were able to understand the questions and also to identify which questions the respondents were reluctant to answer [33].

Those involved in the pre-test were no longer eligible for inclusion in the final survey sample[34].Appropriate revision of the questions was made and the final draft of the questionnaire was administered for the study. A link to the survey was then sent to participants via email. This study utilised purposive sampling which is a type of nonprobability sampling to obtain participants for this study. Purposive sampling involved the researcher making a conscious decision about which individuals would best provide the desired information required for this study [35-36].There is also a snowballing effect taking place as the participants invited others to the survey.

A total of 400 participants agreed to participate in the survey; owing to time constraints a larger sample size could not be obtained. Out of the initial 400 invites sent out only 245 responses were received (a 60% response rate). 180 of the 245 responses were admissible to the study meaning all the questions were answered as required. The other 65 respondents either exited the survey half way or did not attempt to answer some of the questions. An in-depth semi-structured interview was conducted with officials from the IEC who oversee the running of elections in South Africa. The semi-structured interview included questions about the electoral process, challenges with the current electoral process and how theytackle those challenges, the IEC's knowledge of electronic voting and counting technologies, South Africa's readiness for an e-voting or counting system and what factors the IEC thought could hinder the adoption of any of this systems.

This study used thematic data analysis for analysing the data; Braun and Clarke define thematic analysis as a qualitative analytic method for "identifying, analysing and reporting patterns (themes) within data [37]." It minimally organises and describes your data set in rich detail. However, frequently it goes further than this and interprets various aspects of the research topic". In general, thematic analysis involves the searching across a data set to find repeated patterns of meaning [37]. Furthermore, according to Fereday and Muir-Cochrane, thematic analysis is a form of pattern recognition within the data, where emerging themes become the categories for analysis [38]. After going through several literatures on analytical techniques, the study chose to use thematic analysis because of some of the advantages it presents; "Can carefully summarise Key features of a large body of data and offer a thick description of the data set and can highlight similarities and difference across the data set [37]." The DoI framework was also used as a theoretical lens for analysis to validate the analysis process. The themes that emerged from the data were categorised based on the theoretical constructs which are relative advantage, compatibility, and complexity.

## 6. Results/ findings

Figure 1 below represents the percentage of findings from the closed-ended questions from the survey. The questions represent a comparison between electronic voting technologies and the manual paper based system currently used.

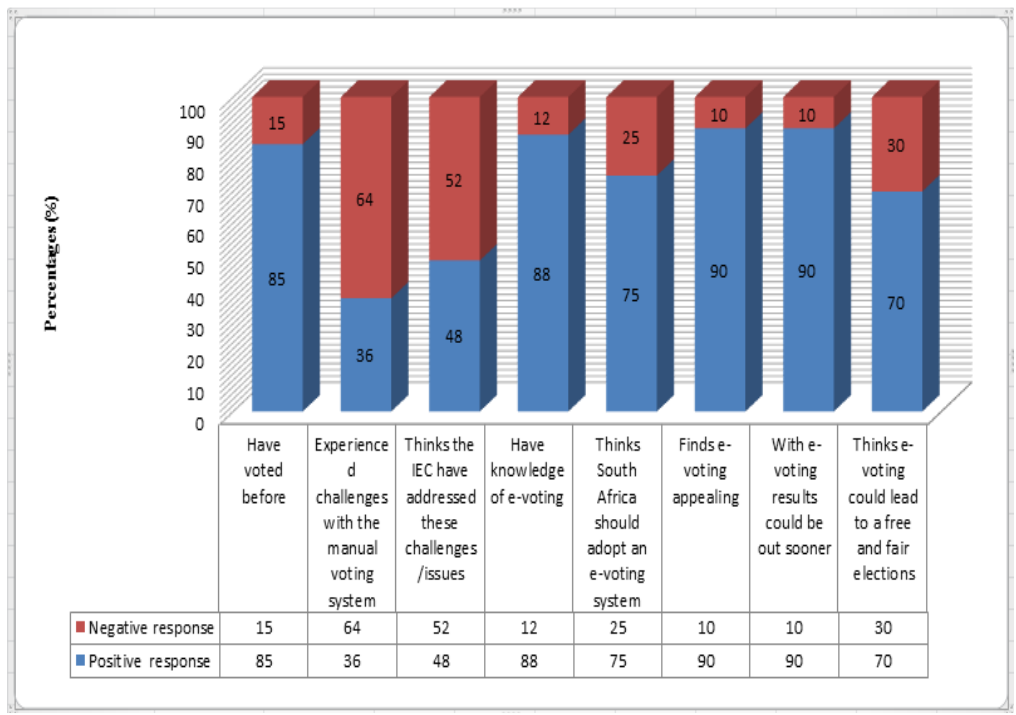

Figure1. Percentage of Responses comparing e-voting and manual paper based voting.

Of the 180 admissible survey responses received, 58% of the participants were females and 42% of the participants were males. 76% of the participants reported their age as between 18-35 years old, while 24% reported being 40 years of age or older. The average age of the respondents was estimated to be 25 years. Ethnicity was found not to be a significant demographic factor therefore; it was not included as a demographic variable in this study.

The findings from the open-ended questions were analysed and then categorised into themes. One of the themes to emerge from the survey data was the *perception of the technology*; it is derived from the category of convenience and accessibility and several other sub-categories. The summarised sub-categories revealed the general voters' perception of electronic voting technology. Statements from some of the responses included the following: "…e-voting technologies would make voting fast and easy therefore reducing the long queues at polling stations." and "…e-voting technology will reduce the amount of money used for running the elections because there will be no printing of paper ballots, etc." This theme fits well within the relative advantages construct context. Based on the findings the participants perceive that an electronic voting system could be better than the current paper based electoral system. *Attitude towards technology* is another theme that emerged from the survey data. According to Rogers attitude towards an innovation is a critical intervening variable in the innovation adoption decision, thus attitude towards a specific information technology (in this case electronic voting technology) is conceptualised as a potential user's assessment of the desirability of using that technology [23].

Based on the responses from the survey, those who had positive perception of electronic voting technology had a positive attitude towards it. Some of the responses included "…with e-voting voters would be able to vote from anywhere no matter what their geographical location is." On the flip side there are those participants who had a negative attitude based on their perception that electronic voting would bring with it a lot of risks such as system hacking, etc.

The concept of compatibility in this study captures a citizen's perception of the similarity of e-voting with other technology innovation accessible to the participants. Responses from the participants suggest that voters who have used systems like SARS e-filling and online banking transactions are likely to view e-voting as consistent with the way they interact with other entities: people, organizations, and government.

Asked if electronic voting would appeal to them compared to the current paper based electoral process, statement of some of the responses included the following:"…there are other electronic services that are offered in the country for example online banking, filling of taxes online, etc. … I think e-voting would appeal to me coz it fits right with such services." The theme attitude towards technology also fits with the DoI construct compatibility. The attitude (negative or positive) that the participants have towards existing innovations could also affect their attitude towards an electronic voting system.

The findings from the interview with the IEC officials revealed several themes. These themes were then analysed to try and fit the themes into the predefined DoI constructs used in this study.Based on the findings the first theme to emerge was *perception of technology;* the IEC officials had the perception that an electronic voting system could be very useful especially when it came to the counting of votes. Asked if South Africa needed an electronic voting system to mitigate some of the challenges the current paper based system was facing, the following are some of the responses: "I think South Africa could use an electronic voting technology that could count the votes as soon as they are cast especially in the under resourced communities." The second theme is the *attitude towards technology;* this mainly comes from the perception of electronic voting by the IEC officials. Even though the IEC perceives the technology to be better than the current paper-based system especially when it comes to the counting of votes, they had a negative attitude towards it citing issues like trust and security risks that could emerge with the introduction of electronic voting technology. There is also the issue of their awareness of other countries that have abandoned the use of electronic voting systems and are back to using the paper based system or a combination of both systems because of the risks that e-voting presents. Some of the responses from the IEC interview include; "…we aware of the countries like the Netherlands and the UK who previously had e-voting systems but now are going back to the manual based system or a combination of both systems… we must first look at the reason these countries are backtracking on e-voting before we think of adopting it."

The construct compatibility was looked at from an angle of how the IEC perceived electronic voting technology to be compatible with theirs and the voters' needs, experience and existing values. Some of the responses from the interview include the following; the IEC states that "… we need a system that can count ballot papers as the votes are cast… " they are also of the view that IEC could use a system to help them run the elections smoothly especially in the under resourced communities. Based on these responses it can be said that the IEC perceive electronic voting technology to be consistent with their need to adopt it. From the findings of the interview the IEC's perception that electronic voting technology could be difficult to use and understand especially for the citizens in the informal settlements whose level of education is low. "… the level of literacy in the informal settlements is very low … some of these voters may find it difficult to use or such a system…"

Another theme from the interview is *infrastructure and resources*; the IEC officials are of the view that even though it would be a good system to have the lack of proper infrastructure and low resources to support the implementation of the system especially in the informal settlements and rural communities will make adopting this technology difficult.

In general the findings reveal that the overall perception of the voters is that e-voting could possibly be a much better system than the current manual paper based system. Voters are of the view that e-voting could make the way in which they cast their votes much easier. The IEC on the other side is of the view that although e-voting as a good system, they also think that the risks and challenges of adopting and implementing such a technology are high. The IEC is also of the view that several factors and have to be considered before e-voting can be adopted in South Africa.

## 7. Discussion

This research was dedicated to studying the adoption of e-voting technologies by exploring how it would diffuse within the South African context using three constructs from Rogers DoI framework. The research also explores the possible factors that could influence the adoption process from the perspective of the voters and the IEC. The study reveals that the three constructs, would exert an important influence on both the IEC and voter's intention to adopt e-voting. The factors that emerged from the findings included the following; Usefulness of the technology, Ease of use, Trust in the technology, resources and infrastructure, and environment.These factors are based on the themes that emerged from the data of both the questionnaire and the interview.

**Usefulness of e-voting technologies**: the findings revealed that the participants favoured electronic voting technology over the current paper based system because of their perception of its usefulness which included convenience of access, time saving, cost (transportation) and the effort it would take to vote. They also favoured electronic voting because they perceived that this system would be able to reduce human error in the electoral process and also increase transparency in the elections. The participants also had the view that such a technology would prevent them from encountering nerve-wracking situations such waiting in long queues at voting stations, intimidation from party agents and many more. These findings confirm the results of the construct relative advantage. The findings from the survey data suggests that the adoption of electronic voting technology is likely possible if they perceive it to be better than the current paper- based system. The IEC also perceived the usefulness of e-voting in terms counting the votes.

International Journal of Managing Information Technology (IJMIT) Vol.5, No.4, November 2013

**Ease of use:** This factor can be compared with the DoI construct complexity. When a technology is perceived by potential adopters as being relatively difficult to use and understand. Based on the findings some participants thought electronic voting system would be difficult to use or understand especially amongst elderly citizens who have no knowledge of such a technology. There was also a concern from the IEC regarding the illiterate citizens in informal settlement who have not had prior use or experience of such technologies might find e-voting difficult to use.

**Trust in the innovation:** The findings show that trust in the technology is a likely factor that could influence the adoption of electronic voting. The participants thought security and privacy issues were factors that might prevent them from trusting and therefore adopting electronic voting technologies. Based on their knowledge or experience of other electronic systems that have been affected by security and privacy issue, these participants thought that if e-voting were not secure enough, their voting right could be under threat and their voting information altered or misused by hackers.

**Resource and infrastructure**: The provision of resources and infrastructure to facilitate the implementation of any innovation is of importance and could influence the adoption of a technology in this case electronic voting. Increased resources would be needed to either provide additional staff training or funding to administer the new voting channel. The findings from the interview with the IEC revealed that the availability of ICTs infrastructure and resources especially in the informal settlements is a factor that could influence their decision where or not to adopt electronic voting. Lack of infrastructure and resources would hinder the adoption of e-voting. Finally, the **Environment** factor, findings from the interview also revealed that the environment within which the potential innovation is to be introduced could influence the adoption of that technology. The interview data showed that before the IEC can decide on adopting an electronic voting technology, they should put into consideration the political environment and the citizens' environment as well. The extent to which both these environments accept the technology is crucial in the IEC's decision on whether to implement electronic voting or not.

## 8. Conclusion

In conclusion the findings in this study revealed the importance of the three DoI constructs in the adoption and diffusion process of e-voting within the South African context. The study also reveals some of the factors that could influence the adoption of e-voting technologies from the perspective of both the voters and the Independent Electoral Commission officials. These factors should be given much consideration before the adoption of e-voting technologies can be considered.

This research has revealed that although e-voting has many potential benefits over the manual voting system, there should be careful deliberation by the decision makers (IEC). The IEC must take into consideration all the factors that could influence the voters both positively and negatively into consideration. This study is of the view that should the ICE successfully explore and address all the factors regarding the adoption of e-voting then South Africa would be able to leverage on the opportunities that e-voting technologies present.

## 10.Limitations to the study

The aspects of this research may limit the interpretation of the results. First, the data was collected solely in the Cape Town area, which may not be representative of the South African population. Secondly the sample size limits the generalizability of the findings of this study to the entire South African concept.The study also focuses on the voters who have access to the internet. The study however gives insight into the factors that could influence the adoption of electronic voting technology within a South African context.

**Authors**

**Principal Author**: Mourine Achieng holds a BTech degree Information Technology: Software Development from the Walter Sisulu University of Technology and a ND: Information Technology from Durban University of Technology and currently a doing MTech Degree with Cape Peninsula University of Technology.

**Co-author:**Dr Ephias Ruhode presently a lecturer at CapePeninsula University of Technology.